\documentclass{llncs}
\usepackage{epsf}

\usepackage{hyperref}
\usepackage{textcomp}
\usepackage{graphicx}
\usepackage{amsmath}
\usepackage{caption}
\usepackage{subcaption}
\usepackage{tabularx}
\usepackage{multirow}

\makeatletter
\newcommand{\printfnsymbol}[1]{%
  \textsuperscript{\@fnsymbol{#1}}%
}
\makeatother

\begin{document}

\title{Identifying short-term interests from mobile app adoption pattern} 

\author{
Bharat Gaind\thanks{Both have contributed equally}, Nitish Varshney\printfnsymbol{1}, Shubham Goel and Akash Mondal}
\institute{Samsung Research Institute Bangalore\\
           Bangalore, India, 560037\\
           \{bharat.gaind, nitish.var, shubham.goel,  a.mondal\}@samsung.com}
\maketitle

\begin{abstract}
With the increase in an average user\textquotesingle s dependence on their mobile devices, the reliance on collecting his browsing history from mobile browsers has also increased. This browsing history is highly utilized in the advertising industry for providing targeted ads in the purview of inferring his short-term interests and pushing relevant ads. However, the major limitation of such an extraction from mobile browsers is that they reset when the browser is closed or when the device is shut down/restarted; thus rendering existing methods to identify the user's short-term interests on mobile devices users, ineffective. In th\texttt{•}is paper, we propose an alternative method to identify such short-term interests by analysing their mobile app adoption (installation/uninstallation) patterns over a period of time. Such a method can be highly effective in pinpointing the user\textquotesingle s ephemeral inclinations like buying/renting an apartment, buying/selling a car or a sudden increased interest in shopping (possibly due to a recent salary bonus, he received). Subsequently, these derived interests are also used for targeted experiments. Our experiments result in up to 93.68\% higher click-through rate in comparison to the ads shown without any user-interest knowledge. Also, up to 51\% higher revenue in the long term is expected as a result of the application of our proposed algorithm. \footnote[1]{This paper has been accepted and presented in the 20th International Conference on Computational Linguistics and Intelligent Text Processing, France, 2019 and will soon be published in the Computaci\'on y Sistemas (Scopus-indexed) journal.}

\end{abstract}

\section{Introduction}
A decade-old advancement in the field of smartphones has made smartphones handy for every user. With the increase in an average user\textquotesingle s dependence on their mobile devices, his dependence on traditional websites, typically browsed via personal computers, is decreasing. Consequently, the focus of the advertising industry has also started to migrate from the web towards mobile devices. However, existing methodologies like cookies, which are used to provide a personalized experience or push targeted advertisements, are not applicable to the mobile ecosystem. The reason is that these cookies are reset, every time the browser is closed or when the phone is shut down/restarted. Additionally, they cannot be shared across apps. Therefore, such traditional methods make mobile marketers handicapped in knowing the user\textquotesingle s interests.

To tackle this, the mobile advertisement industry, these days, is relying on Data Management Platform (DMP), which collects all of the user\textquotesingle s cookies or events coming as first/second/third party data and infers the short-term interests of users. These interests are then used by Demand Side Platforms (DSP’s) or advertising agencies to push targeted ads (formally known as creatives). However, not every advertiser agency or DSP has the luxury of using such DMPs, primarily due to cost restrictions.

In this paper, we propose an alternative method to identify the short-term interests of a user by carefully analysing their mobile app adoption (installation/uninstallation) patterns over a period of time and subsequently, supply these extracted interests to DSPs, that use them to suggest relevant ads for the user. This significantly increases the CTR (Click-Through Rate) of the suggested ads. Our method relies on the key realization that users constantly install and remove mobile-apps on need basis \cite{chittaranjan2013mining} \cite{frey2015reality}, making this data stream valuable and a potential source to deduce their short-term needs, interests or inclinations. Our system first collects the installed-apps list of a mobile-user in a repetitive interval and then extracts installation/uninstallation patterns from it. This installed-apps list contains the applications downloaded and installed by the user, some of which may be pre-installed by the device manufacturer. Features like app-description are then utilized to find the user\textquotesingle s interests using various natural language processing (NLP) based unsupervised methods.


A shortcoming of such an approach is that periodic access to a user's installed apps might seem invasive to an individual's privacy. To avoid such an invasion, we have taken a number of measures. These measures include anonymizing user-identifiers end-to-end and collecting data of only those users who have given consent for data collection and analysis, which they often do to avail intelligent services offering enhanced features. We, therefore, answer all privacy-related questions raised by \cite{jedrzejczyk2009know} and address this issue.

The rest of the paper is organized as follows. In Section 2, we discuss the related works and advancements that have been done in this area. We discuss both the relevant techniques as well as the commercial applications in this field. In Section 3, we give an overview of our system architecture, the modules and their functions. In Section 4, we give a detailed description of the datasets used, followed by results and discussions. In Section 5, we show how our proposed methodology would impact the revenue and CTR of an organization, if applied in the real world.


\section{Related Work}

Identifying a user's interests helps the advertising industry in pushing creatives that align with his preferences. Further, identifying his short-term interests allows the advertising industry to understand when and where the audience wants to engage, the content with which they want to interact, and then select the right place at the right time to push these creatives. This ultimately boosts the Click Through Rates (CTR) and Conversion Rate (CR) of the creatives displayed to the user \cite{merisavo2006effectiveness}.

In previous works, there are predominantly two concepts employed to identify a user\textquotesingle s short-term interests. The first one explores recommendation based on search keywords, which are collected using cookies or events coming as first/\\second/third part-data and are used to infer the short-term interests of users \cite{elmeleegy2013overview}. The main limitation of cookies on mobile browsers is that they reset when the browser is closed or when the phone is shutdown / restarted \cite{WinNT}. Also, as discussed before, they cannot be shared across apps. Furthermore, such approaches cannot be applied to new users or users who like browsing in incognito mode, as such users data is not available with the marketers.

Another concept that is related to identifying the user\textquotesingle s short-term interests is using a snapshot of his installed apps. To the best of our knowledge, research on identifying short-term interests from mobile app adoption pattern has not been conducted. However, some research on predicting user traits \cite{seneviratne2014predicting} \cite{seneviratne2015your} like gender, language, country, religion etc., has been done. These works, typically take a single snapshot of installed apps as input and use supervised learning to categorize users into their traits. Similarly, numerous studies have tried to find a user\textquotesingle s traits, such as, whether the user is single, a parent, his mother tongue, the next app that he is going to install and his life-events \cite{baeza2015predicting} \cite{frey2015reality}. These studies predict a static user property (his traits). Recommending system models using only such user traits to identify user interests have lesser accuracy in comparison to the models using search keywords (discussed in the previous paragraph) \cite{burke2007hybrid}. Thus, existing methods that identifies user traits cannot be directly applied to identify the user\textquotesingle s short-term interests.

Some of the commercial solutions related to identifying a user\textquotesingle s interests are provided by Lotame \footnote{https://www.lotame.com/}, Oracle BlueKai DMP \footnote{https://www.oracle.com/marketingcloud/products/data-management-platform/index.html}, Adobe Audience Manager \footnote{https://www.adobe.com/in/analytics/audience-manager.html} and Salesforce DMP (formerly Krux) \footnote{https://www.salesforce.com/}. These solutions typically organize the marketers' audience data into categories and taxonomies of user-interests and segment the audience to generate various insights.

\begin{figure*}[t]
\label{fig:event_prediction}
\center
  \includegraphics[height=7cm]{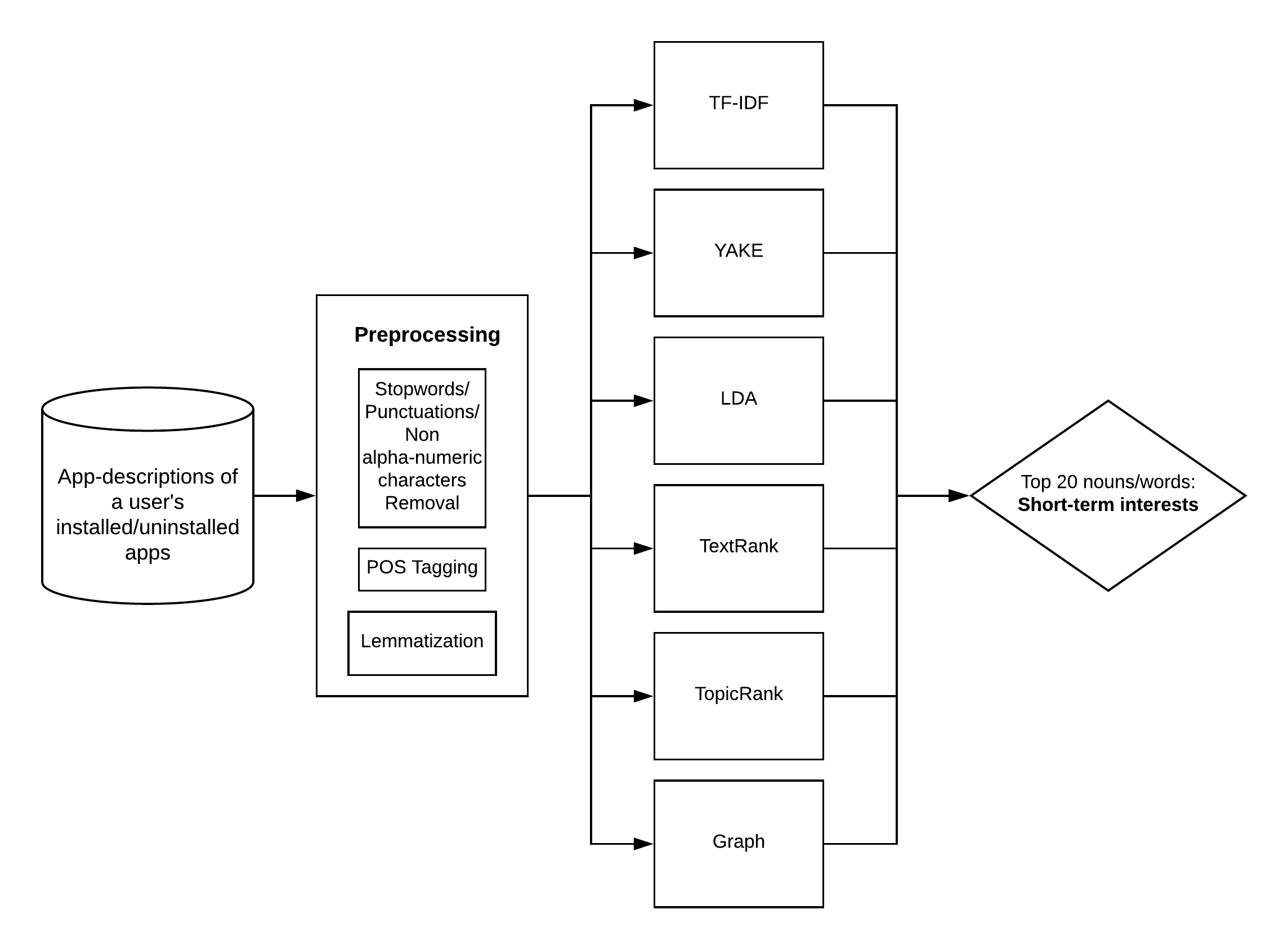}
  \caption{Steps involved in Short-term interest identification}
  \label{fig:step1}
\end{figure*}

\section{Design and Algorithm}

In this section, we explain how we calculate the short-term interests of a user, using his installed/uninstalled apps. Our aim is to retrieve semantically and linguistically important words related to a given app using the app's description (referred to as \textit{app-description} from here on) available from google play store. Our approach consists of various methodologies, all of which are used independently to extract the most important words out of text. We use TF-IDF (Term Frequency-Inverse Document Frequency), YAKE (Yet Another Keyword Extractor), LDA (Latent Dirichlet Allocation), TextRank, TopicRank and Graph methods to achieve this, as shown in Fig. 1. We associate these important words, extracted out of the apps that the user installed over a period of time, as his short-term interests, over that period of time. Similarly, the important words extracted out of the apps that the user uninstalled are considered as events that the user is no longer interested in. All our methodologies are implemented in Python, using various inbuilt and external tools and libraries.

\subsection{Preprocessing}

Preprocessing an app's description (mostly in the form of a paragraph, comprising of multiple sentences) consists of various steps. First, we remove stopwords using \textit{nltk \footnote{Natural Language Toolkit [http://www.nltk.org/], Version 3.3} corpus}. Next, we remove punctuations with the help of \textit{string punctuations}. The third step is removing non-alpha-numeric characters (including non-English languages) using regex. We also find the POS (Parts Of Speech) tags of each of the words in every sentence in the original paragraph using \textit{nltk}, and store these tags in a dictionary, for later use.  Finally, we remove words, that occur very frequently across all app-descriptions using IDF (Inverse Document Frequency). The threshold used for this step is 5\%. In other words, all words that occur at least once in more than 5\% of a randomly chosen app-descriptions set (of size 1000) are removed in this step. Also, before finding their frequency across all app-descriptions, we lemmatize all words in the app-description. This is done using \textit{nltk's WordNetLemmatizer}.

\subsection{Interest identification using TF-IDF}

The first methodology, we use is the standard TF-IDF (Term Frequency-Inverse Document Frequency) technique. It is used to find out the importance of each word in a collection of documents. After preprocessing, as explained above, we calculate TF by finding the frequency of each word and dividing it by the total number of words in the app-description. IDF is calculated by taking the logarithm of a fraction, whose numerator is the frequency of a word, across all app-descriptions, and the denominator is the total number of app-descriptions. Finally, the TF-IDF metric is calculated by multiplying the TF and IDF values. The output of this step is a metric of the importance of each word in the app-description. We then use the stored POS tags, calculated in the previous step, to filter out the nouns in the app-description and output the ones with the highest TF-IDF values.

\subsection{Interest identification using YAKE}

The next approach uses Yet Another Keyword Extractor (YAKE) \cite{yake}, which is a statistical method for multi-lingual keyphrase extraction. Being an unsupervised method, YAKE avoids the problem of the long training process of other supervised methods and does not depend on any dictionaries for topic extraction. We implement the YAKE algorithm using the \textit{pke} library \cite{pke} in python. After preprocessing the input text using the techniques mentioned in section 3.1, we load it as a document in a \textit{pke} YAKE extractor instance. After this, we select 1-3 grams as keyphrase candidates and remove the candidates terminating with a stopword on either side. Next, the candidates are scored using various features extracted from the cleaned input such as word position (since most important words are usually in the beginning), word frequency (the more the word occurs, the better) and other useful features defined in the YAKE algorithm. Finally, we choose 20 words with the lowest scores since they are the most important ones.

\subsection{Interest identification using LDA}

In this approach, we use Latent Dirichlet Allocation (LDA), which is one of the most popular topic-modelling techniques and exploits the fact that every document is originally based on a combination of topics. It tries to backtrack and extract these constituent topics and also outputs a list of relevant words, corresponding to each topic. We use the \textit{gensim} \cite{rehureklrec} library in python to implement LDA. First, we break down the app-description into a list of sentences, and then correspondingly, each sentence into a list of words. We use this corpus as input to gensim to create a dictionary of words, and subsequently, a document-term matrix, which is then fed to the \textit{LdaModel} to find topics. The number of topics selected is 1 (as usually each sentence in app description talks about 1 topic) and the number of passes is 50 (as app-description is usually small, therefore more passes were required). The output of this step is a list of 100 most important words in the app-description, sorted by their degree of associativity to the topic predicted. We then filter the top 20 nouns from these words (picked on the basis of their importance), using the POS tags stored in the preprocessing step and output them as the most important topics in the passage.

\subsection{Interest identification using TextRank}
In this approach, we use TextRank \cite{textrank}, which is a graph-based method for keyword extraction. It implements the idea of voting of a vertex by its adjacent vertices. The vertex having the highest number of votes is the most significant vertex. The score associated with a vertex is calculated using its own number of votes, and the score assigned to the vertices casting these votes. Again, we use the \textit{pke} library in Python to implement this model. In this implementation, nodes are words of certain part-of-speech (nouns and adjectives) and edges represent the distance between word occurrences. Nodes are then ranked by the TextRank graph-based ranking algorithm. A window size of 3 is chosen and the top 5\% vertices are used for phrase generation. Finally, the top 20 words with the highest weights are selected and output as the most important words in the text.

\subsection{Interest identification using TopicRank}

In this approach, we employ TopicRank \cite{topicrank}, which is an unsupervised and a graph-based keyphrase extraction method. Unlike TextRank, this method generates a graph of topics instead of words. Each topic is a collection of similar single and multi-word expressions. The advantage of generating a graph of topics is that the semantic relations between topics are better captured and the graph is more concise. To implement this method, we first preprocess the input data and load it into the \textit{pke} library \cite{pke} TopicRank instance. Then, we select the important topics as groups of similar noun phrases and adjectives in the document using the inbuilt candidate selection method. Next, candidate ranking is done using random walk algorithm. Finally, the top 20 words with the highest weights are output.

\subsection{Interest identification using Graph}

The final approach involves constructing a word co-occurrence network (or graph) using the app-description. Co-occurrence networks/graphs are the collective interconnections of terms based on their paired presence within a specified unit of text. Co-occurrence networks are generated by connecting pairs of terms using a set of criteria defining co-occurrence. Hence, the terms are nodes on the graph and the edges are co-occurrence between those terms. For us, it will be an occurrence in a similar sentence. The system goes through two phases for generating a co-occurrence network: the Graph construction phase and selecting the top 20 nodes from the graph constructed.

In the graph construction phase, first, we split the app-description into sentences. For each sentence, we define co-occurrence between words, which corresponds to an edge between the co-occurrence words. We defined three types of co-occurrence between a pair of words. According to the first type, two words co-occur if they occur in the same sentence. The second type implies co-occurrence when two words occur in the same sentence and are neighbors. Lastly, the third co-occurrence means two words occurring in the same sentence and being at a max distance of two. For example, in the sentence ``stock market is booming high'', $<$ ``stock'', ``high'' $>$ and $<$ ``stock'', ``booming'' $>$ co-occur by definition 1. Also, $<$ ``stock'', ``market'' $>$ co-occur by definition 2 and 3. In order to realize all such possibilities, we generated three separate graphs based on different co-occurrences. We also removed all self-loops present in the graph.

Next, we consider words (nodes) from our co-occurrence graph using various graph metrics. Various graph metrics considered for experimentation are \textit{Degree}, \textit{Page Rank}, \textit{Betweenness centrality} and \textit{Closeness} centrality. We took the top 20 words, based on the aforementioned graph metrics. All the graph operations including construction and metric calculations are done using \textit{networkX \cite{networkx}} in python.

\section{Results}

As mentioned before, we take the top 20 most important words as the output of each of the methodologies explained above. In this section, we explain how we evaluate these methodologies. Our project work in the ads domain allows us to access the current apps installed in the user's Android phone. Using this data, we find all installed and uninstalled apps of users during a fixed period of time (July 1 to July 15, 2018, in our case). We then apply each approach explained above to find the 20 most important words out of the app-descriptions of each of these installed/uninstalled apps. These words act as short-term interests (or dislikes, in the case of uninstalls) of the user. Our evaluation methods compute an estimator of the potential increase in CTR that could be generated by the application of this new methodology. The experiments are conducted on a dataset of real users in an offline manner after imitating a real-time environment. During imitation, we have considered all eligible bids (including losing bids in the ad-bidding process) for each advertisement request. Each bid object contains the advertiser's unique identifier, the advertised product description and bidding price.

Now, we need a dataset that acts as ground truth to assess our output. Another dataset available to us, as part of our project, is the CTR-dataset (Click Through Rate-dataset) of around 1 million users. We use it to identify the clicks and the impressions of the ads shown to the user from July 7 to July 21, 2018. The reason this period of time is chosen is that we assume that on an average, the short-term interests of the user in the period July 1 to July 15, will be at their peak on July 7th. Therefore, the probability of him clicking an ad of the similar category (as his installed apps) will be the highest during July 7th to July 21st (chosen because his short-term interests might persist for around 7 more days after he's finished installing/uninstalling apps between July 1st and July 15). Since we consider the event that a user clicks on an ad of a particular category as an indicator of his short-term interests, this dataset can act as ground-truth.

Next, we calculate the CTR ($CTR_1$) for all creatives (or ads) that are similar to his installed/uninstalled apps. This similarity is calculated by finding the top 20 words for both the ads and the installed apps, using the various approaches discussed above, and then using \textit{nltk wordnet} to find synonyms (Two apps are similar if their app-descriptions have at least two common words/synonyms). $CTR_1$ represents what the CTR would have been if the creatives were chosen according to the results predicted by our methodologies. We also calculate CTR ($CTR_2$) of the ads that are dissimilar to the user's installed apps. $CTR_2$ represents the probability of a user clicking an ad if our approach is not used. 

\begin{table}[]
\caption{CTR enhancement for interests identified based on installed apps.}
\label{tab:block_results1}
\centering
\begin{tabular}{llrrrr}
\hline
\bf Method                                                                                                &                & \bf \begin{tabular}[c]{@{}r@{}}CTR1\\ (\%age)\end{tabular} & \bf \begin{tabular}[c]{@{}r@{}}CTR2\\ (\%age)\end{tabular} & \bf \begin{tabular}[c]{@{}r@{}}Increase in CTR \\ (\%age)\end{tabular} & \bf \begin{tabular}[c]{@{}r@{}}Applicable Bids \\ (\%age)\end{tabular} \\ \hline
TF-IDF                                                                                                &                & 4.99                                                   & 3.37                                                   & 48.07                                                              & 61                                                                  \\
\hline
YAKE                                                                                                  &                & 4.09                                                   & 3.57                                                   & 14.57                                                              & 64                                                                  \\
\hline
LDA                                                                                                   &                & 5.28                                                   & 2.79                                                   & \textbf{89.20}                                                     & 40                                                                  \\
\hline
TextRank                                                                                              &                & 5.52                                                   & 2.85                                                   & \textbf{93.68}                                                    & 24.6                                                                \\
\hline
TopicRank                                                                                            &                & 5.34                                                   & 3.28                                                   & \textbf{62.8}                                                      & 64.7                                                               \\
\hline
\multirow{3}{*}{\begin{tabular}[c]{@{}l@{}}Degree \\ graph\\ ranking \end{tabular}}           & Co-occ. type 1 & 4.52                                                   & 3.69                                                   & 22.49                                                              & 78.8                                                                \\
                                                                                                      & Co-occ. type 2 & 4.51                                                   & 3.49                                                   & 29.23                                                              & 78.6                                                                \\
                                                                                                      & Co-occ. type 3 & 4.64                                                   & 3.42                                                   & \textbf{35.67}                                                              & \textbf{80.6}                                                       \\ 
\hline
\multirow{3}{*}{\begin{tabular}[c]{@{}l@{}}PageRank \\ graph\\ ranking \end{tabular}}         & Co-occ. type 1 & 4.38                                                   & 3.87                                                   & 13.18                                                              & 79.9                                                                \\
                                                                                                      & Co-occ. type 2 & 4.51                                                   & 3.50                                                   & \textbf{28.85}                                                              & \textbf{81.3}                                                       \\
                                                                                                      & Co-occ. type 3 & 4.50                                                   & 3.50                                                   & 28.57                                                              & 79.8                                                                \\
\hline
\multirow{3}{*}{\begin{tabular}[c]{@{}l@{}}Betweenness\\ centrality \\ ranking \end{tabular}} & Co-occ. type 1 & 4.61                                                   & 3.29                                                   & 39.74                                                              & 77.6                                                                \\
                                                                                                      & Co-occ. type 2 & 4.71                                                   & 3.26                                                   & \textbf{44.48}                                                              & \textbf{78.8}                                                       \\
                                                                                                      & Co-occ. type 3 & 4.66                                                   & 3.29                                                   & 41.64                                                              & 78.4                                                                \\
\hline
\multirow{3}{*}{\begin{tabular}[c]{@{}l@{}}Closeness \\ centrality \\ ranking \end{tabular}}  & Co-occ. type 1 & 4.52                                                   & 3.79                                                   & 19.26                                                              & 76.6                                                                \\
                                                                                                      & Co-occ. type 2 & 4.52                                                   & 3.86                                                   & 17.09                                                              & 80.3                                                                \\
                                                                                                      & Co-occ. type 3 & 4.71                                                   & 3.31                                                   & \textbf{42.29}                                                              & \textbf{82.1}                                                       \\
\hline
\end{tabular}
\end{table}

\begin{table}[]
\caption{CTR enhancement for interests identified based on uninstalled apps.}
\label{tab:block_results2}
\centering
\begin{tabular}{llrrrr}
\hline
\bf Method                                                                                                &                & \bf \begin{tabular}[c]{@{}r@{}}CTR1\\ (\%age)\end{tabular} & \bf \begin{tabular}[c]{@{}r@{}}CTR2\\ (\%age)\end{tabular} & \bf \begin{tabular}[c]{@{}r@{}}Increase in CTR \\ (\%age)\end{tabular} & \bf \begin{tabular}[c]{@{}r@{}}Applicable Bids \\ (\%age)\end{tabular} \\ \hline
TF-IDF                                                                                                &                & 3.13                                                   & 3.98                                                   & 27.16                                                              & \textbf{75.89}                                                               \\
\hline
YAKE                                                                                                  &                & 3.22                                                   & 4.53                                                   & 40.68 & 72.21                                                               \\
\hline
LDA                                                                                                   &                & 3.54                                                   & 4.76                                                   & 34.46 & \textbf{85.26}                                                               \\
\hline
TextRank                                                                                              &                & 3.29                                                   & 5.15                                                   & 56.53 & 74.00                                                               \\
\hline
TopicRank                                                                                             &                & 3.23                                                   & 4.71                                                   & 45.82                                                              & \textbf{78.84}                                                               \\
\hline
\multirow{3}{*}{\begin{tabular}[c]{@{}l@{}}Degree \\ graph\\ ranking \end{tabular}}         & Co-occ. type 1 & 2.92                                                   & 5.80                                                   & 98.63                                                              & 60.42                                         \\
                                                                                                      & Co-occ. type 2 & 2.95                                                   & 5.83                                                   & 97.62                                                              & 60.63                                                               \\
                                                                                                      & Co-occ. type 3 & 2.91
         & 5.89                                                   & \textbf{102.41}                                                              & \textbf{63.89}     \\
\hline
\multirow{3}{*}{\begin{tabular}[c]{@{}l@{}}PageRank \\ graph\\ ranking \end{tabular}}         & Co-occ. type 1 &                  2.95                                                   & 5.62                                                   & 90.51                                                              & 60.03                                                               \\
                                                                                                      & Co-occ. type 2 & 2.89                                                   & 5.93                                                   & \textbf{105.19}                                                             & \textbf{63.26}                                                               \\
                                                                                                      & Co-occ. type 3 & 2.91                                                   & 5.88                                                  & 102.61                                                             & 66.03  \\
\hline
\multirow{3}{*}{\begin{tabular}[c]{@{}l@{}}Betweenness\\ centrality \\ ranking \end{tabular}} & Co-occ. type 1 & 2.93                                                   & 5.77                                                   & 96.92                                                              & 60.03                                                               \\
                                                                                                      & Co-occ. type 2 & 2.94                                                   & 5.83                                                   & 98.30 & 62.63 \\
                                                                                                      & Co-occ. type 3 & 2.96                                                   & 5.51                                                   & 86.15                                                              & 59.47                                                               \\
\hline
\multirow{3}{*}{\begin{tabular}[c]{@{}l@{}}Closeness \\ centrality \\ ranking \end{tabular}}  & Co-occ. type 1 & 2.98                                                   & 5.97                                                   & 100.33                                                             & 60.63                                                               \\
                                                                                                      & Co-occ. type 2 & 2.87                                                   & 6.13                                                   & \textbf{113.58}                                                             &  \textbf{68.06}                                                              \\
                                                                                                      & Co-occ. type 3 & 2.94                                                   & 6.11                                                   & 107.82                                                             & 60.21 \\                                                              
\hline
\end{tabular}
\end{table}

The evaluation is based on the hypothesis that for a particular user and his short-term interests identified based on installed apps, $CTR_1$ $>$ $CTR_2$ (since if our methodologies are used, the CTR is expected to increase, as the ads now have become more personalized/interesting to the user). Similarly, for interests identified based on uninstalled apps, our hypothesis is $CTR_1$ $<$ $CTR_2$ (since our methodologies are used, in this case, to identify ads that are not of interest to the user anymore). The average $CTR_1$ and $CTR_2$ over 34,908 users for install-based interest identification and 28,226 users for uninstall-based interest identification, computed this way is depicted in Table~\ref{tab:block_results1} and ~\ref{tab:block_results2}. The CTR increase column denotes how much the CTR increased from $CTR_2$ to $CTR_1$ in case of Table~\ref{tab:block_results1}, and $CTR_1$ to $CTR_2$, in case of Table \ref{tab:block_results2}. 

Additionally, our algorithm is not applicable to all the bids. (For some bids, the top keywords identified did not match with any of the user's installed/uninstalled apps' top keywords). Hence, we have also tabulated the percentage of bids (denoted as applicable bids), for which different methods were applicable. As can be observed from Table~\ref{tab:block_results1}, the TextRank model shows the maximum CTR increase of 93.68\% for interests identified based on installed apps, but its applicability is low. Similarly, In Table~\ref{tab:block_results2}, the Closeness Centrality ranking method of type 2 shows the maximum CTR increase of 113.58\% for uninstallation-apps based non-interest identification, with a relatively low applicability. To increase the applicability while maintaining the increase in CTR, we have implemented various priority-based Hybrid models, comprising of 3 models having the maximum increase in CTR and one model having high applicability. In a hybrid model, we try to find the common keywords between a user's identified interests (installs/uninstalls) and the incoming bids, using the the highest-priority model (LDA, for instance, in case of installs). If this model fails, the next model is tried out to find these common words and so on. Table~\ref{tab:hybrid_results1} and \ref{tab:hybrid_results2} denote the various hybrid models implemented. Each model comprises of 4 constituent models as discussed above. The hybrid model, \textit{LDA\_TopicRank\_TextRank\_Degree-3}, for instance, consists of the LDA, TopicRank, TextRank and the degree graph (Co-occurrence type 3) models. A significant improvement can be seen in the applicability of the bids (85.6\% and 82\%,  respectively) in the hybrid models, while maintaining an impressive CTR increase (90.13\% and 121.82\%, respectively), as shown in Table\ref{tab:hybrid_results1} and \ref{tab:hybrid_results2}.

\begin{table}[]
\caption{CTR enhancement for interests identified based on installed apps using priority-based Hybrid models.}
\label{tab:hybrid_results1}
\centering
\begin{tabular}{lrrrr}
\hline
\bf Hybrid Method       & \bf \begin{tabular}[c]{@{}r@{}}CTR1\\ (\%age)\end{tabular} & \bf \begin{tabular}[c]{@{}r@{}}CTR2\\ (\%age)\end{tabular} & \bf \begin{tabular}[c]{@{}r@{}}Increase in \\ CTR (\%age)\end{tabular} & \bf \begin{tabular}[c]{@{}r@{}}Applicable Bids \\ (\%age)\end{tabular} \\ \hline
\begin{tabular}[c]{@{}l@{}}LDA\_TopicRank\_TextRank\_\\ Degree-3\end{tabular}     & 6.822 & 3.674 & 85.65          & 85.6            \\
\begin{tabular}[c]{@{}l@{}}LDA\_TopicRank\_TextRank\_\\ PageRank-2\end{tabular}  & 6.773 & 3.701 & 82.97          & 86.8            \\
\textbf{\begin{tabular}[c]{@{}l@{}}LDA\_TopicRank\_TextRank\_\\ BwCent-2\end{tabular}} & 6.934 & 3.647 & \textbf{90.13} & \textbf{85.6}   \\
\begin{tabular}[c]{@{}l@{}}LDA\_TopicRank\_TextRank\_\\ ClCent-3\end{tabular}  & 6.772 & 3.671 & 84.47          & 85    \\
\hline       
\end{tabular}
\end{table}

\begin{table}[t]
\caption{CTR enhancement for interests identified based on uninstalled apps using priority-based Hybrid models.}
\label{tab:hybrid_results2}
\centering
\begin{tabular}{lrrrr}
\hline
\bf Hybrid Method       & \bf \begin{tabular}[c]{@{}r@{}}CTR1\\ (\%age)\end{tabular} & \bf \begin{tabular}[c]{@{}r@{}}CTR2\\ (\%age)\end{tabular} & \bf \begin{tabular}[c]{@{}r@{}}Increase in \\ CTR (\%age)\end{tabular} & \bf \begin{tabular}[c]{@{}r@{}}Applicable Bids \\ (\%age)\end{tabular} \\ \hline
\begin{tabular}[c]{@{}l@{}}Degree-3\_PageRank-2\_ \\ ClCent-2\_TF-IDF\end{tabular}     & 3.239 & 6.698 & 106.75          & 79.6            \\
\begin{tabular}[c]{@{}l@{}}Degree-3\_PageRank-2\_ \\ ClCent-2\_LDA\end{tabular} & 3.365 & 7.085 & 110.52 & 88.6   \\
\textbf{\begin{tabular}[c]{@{}l@{}}Degree-3\_PageRank-2\_ \\ ClCent-2\_TopicRank\end{tabular}}  & 3.277 & 7.27 & \textbf{121.82}          & \textbf{82}    \\
\hline       
\end{tabular}
\end{table}



\section{Impact on Revenue}
In the advertising industry, there are majorly three players from a pricing perspective. The first ones are the advertisers who want to acquire new users via advertisement. Second, there are publishers who have a dedicated space where the advertisement can be shown to the user, formally known as an ad inventory. And lastly, there are intermediaries (mostly DSPs) making the match via the bidding process. The advertisers usually want to pay for per-user action (a click or an app install), which is formally known as Cost per Action (CPA) and the publishers want to earn per impression, formally known as Cost per Mile (CPM) impression. Hence, the intermediaries also need to perform arbitrage. We define arbitrage as a process of converting CPA to CPM. The prevalent methodologies applied for performing arbitrage predict the CTR for each bid, which can then be used as CPA*CTR to get CPM (assuming intermediary is not taking any cut out of it). For example, consider the case when an advertiser has an ad budget of \$10 per click and the intermediary predicted that the CTR for the advertisement is 0.1 for a user's request. In such a scenario, the intermediary may bid for \$1 (\$10*0.1) for an ad space and the advertisement would be selected for display, if \$1 is the highest-priced valid bid received.

From the last decade, publishers also like to provide targeted advertisements, even at the cost of initial revenue loss, as it improves the long-term value of their brand by not cluttering their ad space. In the process, an initial revenue loss would likely result as the highest-priced bid might not be the most liked bid by the user (identified using our methodology). This initial revenue loss is tabulated in Table \ref{tab:rev_1}, where we have shown the initial revenue impact on publishers after applying various methodologies proposed in this paper. In this table, we have computed the revenue loss on 1 million user requests containing an ad inventory of three different publishers. 
For computing the estimated long-term impact, the increase in CTR (\%age) and applicable users (\%age) metrics have been taken from Table \ref{tab:block_results1}.

\begin{table}[]
\caption{Real time impact on application of our algorithm.}
\label{tab:rev_1}
\centering
\begin{tabular}{llrrrr}
\hline
\bf Method                                                                                               &                & \textbf{\begin{tabular}[c]{@{}r@{}}Initial\\ \hspace{0.7cm}Impact\\ (\%age)\end{tabular}} & \begin{tabular}[c]{@{}r@{}}\hspace{0.3cm}Increase \\in CTR\\ (\%age)\end{tabular} & \begin{tabular}[c]{@{}r@{}}\hspace{0.1cm}Applicable \\ Bids\\ (\%age)\end{tabular} & \textbf{\begin{tabular}[c]{@{}r@{}}Estimated \\Long-term \\\hspace{0.1cm}Impact(\%age)\end{tabular}} \\ \hline
TF-IDF                                                                                               &                & -20.1435                                                                  & 48.07                                                                       & 61                                                                 & 18.60584                                                                               \\ \hline
YAKE                                                                                                 &                & -13.6853                                                                  & 14.57                                                                       & 64                                                                 & 21.83496                                                                               \\ \hline
LDA                                                                                                  &                & -22.5013                                                                  & 89.2                                                                        & 40                                                                 & \textbf{37.21169}                                                                      \\ \hline
TextRank                                                                                             &                & -23.1164                                                                  & 93.68                                                                      & 24.6                                                               & \textbf{51.102}                                                                        \\ \hline
TopicRank                                                                                            &                & -17.9908                                                                  & 62.8                                                                        & 64.7                                                               & \textbf{29.1184}                                                                       \\ \hline
\multirow{3}{*}{\begin{tabular}[c]{@{}l@{}}Degree \\ graph\\ ranking\end{tabular}}                   & Co-occ. type 1 & -13.1727                                                                  & 22.49                                                                       & 78.8                                                               & 7.842132                                                                               \\ 
                                                                                                     & Co-occ. type 2 & -9.48232                                                                  & 29.23                                                                       & 78.6                                                               & 13.12148                                                                               \\ 
                                                                                                     & Co-occ. type 3 & -8.14967                                                                  & 35.67                                                                       & 80.6                                                               & \textbf{17.68324}                                                                      \\ \hline
\multirow{3}{*}{\begin{tabular}[c]{@{}l@{}}PageRank \\ graph\\ ranking\end{tabular}}                 & Co-occ. type 1 & -13.4803                                                                  & 13.18                                                                       & 79.9                                                               & 5.894413                                                                               \\ 
                                                                                                     & Co-occ. type 2 & -9.53357                                                                  & 28.85                                                                       & 81.3                                                               & \textbf{11.48129}                                                                      \\ 
                                                                                                     & Co-occ. type 3 & -9.78985                                                                  & 28.57                                                                       & 79.8                                                               & 11.22501                                                                               \\ \hline
\multirow{3}{*}{\begin{tabular}[c]{@{}l@{}}Betweenness\\ centrality \\ ranking\end{tabular}}         & Co-occ. type 1 & -10.0974                                                                  & 39.74                                                                       & 77.6                                                               & 17.32445                                                                               \\ 
                                                                                                     & Co-occ. type 2 & -15.223                                                                   & 44.48                                                                       & 78.8                                                               & 15.42799                                                                               \\
                                                                                                     & Co-occ. type 3 & -9.53357                                                                  & 41.64                                                                       & 78.4                                                               & \textbf{19.52845}                                                                      \\ \hline
\multirow{3}{*}{\begin{tabular}[c]{@{}l@{}}Closeness \\ centrality \\ ranking \end{tabular}} & Co-occ. type 1 & -13.1727                                                                  & 19.26                                                                       & 76.6                                                               & 9.431061                                                                               \\ 
                                                                                                     & Co-occ. type 2 & -13.3778                                                                  & 17.09                                                                       & 80.3                                                               & 5.996929                                                                               \\
                                                                                                     & Co-occ. type 3 & -11.0712                                                                  & 42.29                                                                       & 82.1                                                               & \textbf{17.99077}      \\ \hline                                                               
\end{tabular}
\end{table}

In Table \ref{tab:rev_2}, through an example, we try to explain how the revenue has increased in the long-term for the publishers in Table \ref{tab:rev_1}. We consider two advertisers, having an ad budget of \$12 and \$10 per click respectively, who want to get 100 clicks each for their advertisements. The intermediary has set the default CTR (to convert CPA to CPM) for both the advertisements as 0.1 (in the absence of any system to identify user interests). Additionally, assume that the publisher has an ad inventory of 1000. Therefore, the publisher will earn \$1200 (=\$1.2 cost per impression * 1000 impressions), when our system is not in place. Now, suppose that the second ad is more aligned to the user's interests (as predicted by our algorithm), as a result of which it is chosen by the publisher instead of the first one. In this case, the publisher will initially earn \$1000, which would mean an initial loss of \$200 for every 1000 impressions. However, in the long term, the second advertiser's target of reaching 100 clicks would be achieved after only 691 impressions (assuming a 44.67\% average increase in CTR and 100\% applicability from Table \ref{tab:rev_1}). In this case, the predicted CTR would be 0.14467, which is why the number of impressions needed would be 100 clicks divided by 0.14467, which comes to approximately 691 impressions. Since the goal of the second advertiser (of achieving 100 clicks) has been achieved in fewer than 1000 impressions, it will not participate anymore in the bidding process, which leaves the ad space for 309 impressions available to the publisher. The publisher can then show the first advertiser's ad in this ad space, which will generate an additional revenue of \$371 (=309 impressions*\$12 per click*0.1 predicted CTR). Hence, the total revenue of \$1000 + \$371 would be earned by the publisher, which is \$171 more than the revenue earned when our system was not deployed. Since, in real-world scenarios, publishers display millions of impressions per month, applying our algorithm could have a significant impact on the long-term revenue of the publishers.


\begin{table}
\caption{Long-term impact on publishers revenue.}
\label{tab:rev_2}
\centering
\subcaption*{Part A : Without application of our algorithm}
\begin{tabular}{lrrr}
\hline
\bf  & \bf Advertiser's ad & \bf Intermediary's & \bf\hspace{0.1cm} Publisher's revenue\\
\bf  & \bf budget per click & \bf\hspace{0.1cm}predicted CTR & \bf per impression\\
\hline
Advertiser1  & \$12    & 0.1    & \$12*0.1=\$1.2    \\
Advertiser2  & \$10    & 0.1    & \$10*0.1=\$1.0   \\
\hline 
\end{tabular}
\bigskip
\subcaption*{Part B : On application of our algorithm on the second advertiser's bid}
\begin{tabular}{lrrr}
\hline
\bf  & \bf Advertiser's ad & \bf Intermediary's & \bf\hspace{0.1cm} Publisher's revenue\\
\bf  & \bf budget per click & \bf\hspace{0.1cm} predicted CTR & \bf per impression\\
\hline
Advertiser1  & \$12    & 0.10000  & \$12*0.10000=\$1.2000 \\
Advertiser2  & \$10    & 0.14467  & \$10*0.14467=\$1.4467  \\
\hline 
\end{tabular}
\end{table}

\section{Conclusion}
In this paper, we propose various methodologies to identify the short-term interests of a user by analysing his mobile app adoption (installation/uninstallation) patterns over a period of time. Such a method can be highly effective in pinpointing the user\textquotesingle s ephemeral inclinations. Our experiments result in around 94\% higher click-through rate (in case of installed-apps based interest identification using TextRank algorithm) and around 113\% higher click-through rate (for uninstalled-apps based non-interest identification using the closeness-centrality ranking method of graphs), in comparison to the ads shown without any user-interest knowledge. Further, we implement several hybrid models having both a high CTR increase and bid-applicability (as high as 121.82\% and 82\%, in the dislike-identification case). Also, around 51\% higher revenue in the long-term is expected as a result of the application of our proposed algorithm. In future, we would optimize our methodologies to decrease their execution-times, since our priority in this paper was achieving a higher CTR increase. Also, we would work on unifying various installation and uninstallation-based models to make the overall system better personalized to the user's interests.

\bibliographystyle{splncs}
\bibliography{paper}
\end{document}